\documentclass[reprint,superscriptaddress,aps,prx]{revtex4-2}

\usepackage[usenames,dvipsnames]{xcolor}
\usepackage[utf8]{inputenc}
\usepackage{bm}
\usepackage{amsfonts}
\usepackage{graphicx,psfrag}
\usepackage{booktabs}
\usepackage{amsmath}
\usepackage{mathtools}
\usepackage{physics}
\usepackage{calc}
\usepackage{makecell}
\usepackage{enumitem}
\usepackage{float}
\usepackage{natbib}
\usepackage{tikz}
\usepackage{soul,xcolor}
\usepackage[colorlinks=true,linkcolor=Blue,citecolor=Blue,urlcolor=Blue]{hyperref}

\usepackage{xfp}

\usepackage[caption=false]{subfig}
\usepackage[capitalise]{cleveref}

\DeclarePairedDelimiter\floor{\lfloor}{\rfloor}

\newcommand{\rev}{\color{black}}
\newcommand{\revise}{\color{black}}
\newcommand{\new}{\color{black}}

\begin{document}

\title{Deeper but smaller:\\ Higher-order interactions increase linear stability but shrink basins}

\author{Yuanzhao Zhang}
\email{yzhang@santafe.edu}
\affiliation{Santa Fe Institute, Santa Fe, NM 87501, USA}

\author{Per Sebastian Skardal}
\affiliation{Department of Mathematics, Trinity College, Hartford, CT 06106, USA}

\author{Federico Battiston}
\affiliation{Department of Network and Data Science, Central European University, 1100 Vienna, Austria}

\author{Giovanni Petri}
\affiliation{NP Lab, Network Science Institute, Northeastern University London, London, UK}
\affiliation{Department of Physics, Northeastern University, Boston, MA 02115, USA}
\affiliation{CENTAI Institute, 10138 Torino, Italy}

\author{Maxime Lucas}
\email{maxime.lucas.work@gmail.com}
\affiliation{CENTAI Institute, 10138 Torino, Italy}

\begin{abstract} 
A key challenge of nonlinear dynamics and network science is to understand how higher-order interactions influence collective dynamics. Although many studies have approached this question through linear stability analysis, less is known about how higher-order interactions shape the global organization of different states. Here, we shed light on this issue by analyzing the rich patterns supported by identical Kuramoto oscillators on hypergraphs. We show that higher-order interactions can have opposite effects on linear stability and basin stability: they stabilize twisted states (including full synchrony) by improving their linear stability, but also make them hard to find by dramatically reducing their basin size. Our results highlight the importance of understanding higher-order interactions from both local and global perspectives.
\end{abstract}

\maketitle

\section{Introduction}

{\rev Higher-order interactions are couplings that connect more than two units simultaneously and in a nonlinear way so that it cannot be decomposed into a linear combination of pairwise interactions} \cite{lambiotte2019networks,battiston2020networks,torres2021and,battiston2021physics,battiston2022higher,bick2023higher}.
Such nonpairwise interactions are crucial in shaping complex dynamical processes such as contagion and cooperation in social networks \cite{iacopini2019simplicial,schaub2020random,lucas2023simplicially,landry2020effect,st2021universal, alvarez2021evolutionary,civilini2024explosive}, information processing in the brain \cite{petri2014homological,giusti2016two,parastesh2022synchronization,santoro2023higher,santoro2024higher}, and synchronization in coupled oscillators \cite{leon2019phase,matheny2019exotic,gengel2020high,topal2023reconstructing,leon2024higher}.
Understanding how they influence collective dynamics is thus essential. 
A variety of studies have approached this challenge from a linear stability perspective, which characterizes how states such as synchronization and consensus respond to small perturbations
\cite{lucas2020multiorder,millan2020explosive,neuhauser2020multibody,skardal2021higher,zhang2021unified,salova2021cluster,ferrazdearruda2021phase,gambuzza2021stability,gallo2022synchronization,zhang2023higher, carletti2023globala, nurisso2024unified}.
However, little attention has been paid to basin stability \cite{menck2013basin}, a global measure based on the size of basins of attraction, which dictates the system's response to large perturbations \cite{milnor1985concept,ott2002chaos,aguirre2009fractal,menck2014dead,zhang2020critical}.

In this paper, we provide a more complete picture of how higher-order interactions influence dynamical patterns, in terms of both linear and basin stability. 
We show that higher-order interactions can have opposite effects: Such interactions can increase the number of ordered states by making them linearly stable; at the same time, higher-order interactions also dramatically shrink their attraction basins, effectively hiding them from detection.
As a result, states such as full synchrony may be stable but unreachable from random initial conditions.

To demonstrate this point, we consider $n$ identical phase oscillators coupled through both pairwise and triadic interactions:
\begin{equation}
\begin{split}
    \dot{\theta}_i = \omega & + \frac{\sigma}{k_i^{(1)} } \sum_{j=1}^{n} A_{ij} \sin(\theta_j-\theta_i) \\
    & + \frac{\sigma_\Delta}{2 k_i^{(2)}} \sum_{j,k=1}^{n} B_{ijk} \sin(\theta_j+\theta_k-2\theta_i). 
\end{split}
\label{eq:kuramoto}
\end{equation}
\Cref{eq:kuramoto} is a generalization of the Kuramoto model \cite{strogatz2000kuramoto}, {\new which can be derived exactly from the phase reduction of weakly coupled, nearly identical limit-cycle oscillators \cite{leon2024higher}.
In this sense, the Kuramoto dynamics represent a canonical model for a broad class of real-world systems exhibiting periodic oscillations.
For example, Kuramoto dynamics with higher-order interactions have been used to analyze the collective dynamics of nanoelectromechanical oscillators observed in experiments \cite{matheny2019exotic}.}
Here, $\theta_i \in S^1$ represents the phase of oscillator $i$ and $\omega$ is their common frequency.
The adjacency tensors determine which oscillators interact: $A_{ij}=1$ if nodes $i$ and $j$ have a pairwise connection, and zero otherwise. 
Similarly, $B_{ijk}=1$ if and only if nodes $i$, $j$, and $k$ are coupled through a triadic interaction. 
The coupling strengths are given by $\sigma$ and $\sigma_\Delta$, respectively, and are normalized by $k_i^{(\ell)}$, the $\ell$th order degree of node $i$.

The case of pairwise coupling ($\sigma_\Delta=0$) has been studied in detail from both linear and basin stability perspectives. 
While full synchrony $\theta_i(t)=\theta_j(t)\;\forall\, i,j,t$ is always an attractor of \cref{eq:kuramoto}, additional attractors can emerge when networks are not too dense \cite{townsend2020dense,kassabov2021sufficiently,kassabov2022global,abdalla2022expander}.
For ring networks, these attractors are \textit{twisted states} and they emerge for link density below $0.68$ \cite{wiley2006size}.
In a $q$-twisted state $\bm{\theta}^{(q)}$, the phases make $q$ full twists around the ring and satisfy $\theta_k^{(q)} = 2 \pi k q/n + C,$  where $q$ is the \textit{winding number} [c.f. \cref{fig:linear_stability}(a)]. In particular, the fully synchronized state corresponds to $q=0$.
For rings with nearest-neighbor couplings,
the number of attractors grows linearly with $n$, since twisted states with up to $n/4$ twists are stable \cite{delabays2016multistability,manik2017cycle}.
A fruitful line of research aims to characterize the basins of the coexisting twisted states, which has revealed interesting scaling relations between basin size and winding number \cite{wiley2006size,delabays2017size,zhang2021basins} as well as tentacle-like structures in the basins \cite{ashwin2012calculations,martiniani2016structural,zhang2021basins}.

\section{Results}

\subsection{Linear stability analysis}

\begin{figure*}[t]
\centering
\includegraphics[width=.95\linewidth]{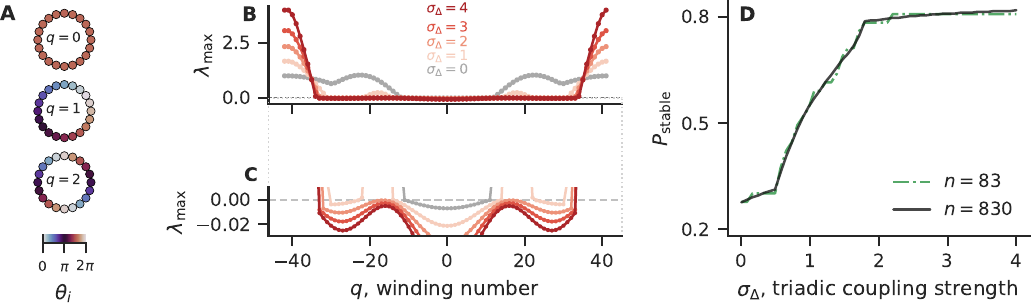}
\caption{
\textbf{Higher-order interactions improve the linear stability of twisted states.} 
(a) Example twisted states with different winding numbers, for $n=20$. Full synchrony corresponds to $q=0$.
(b) Linear stability (measured by the largest transverse Lyapunov exponent $\lambda_{\max}$) of $q$-twisted states, for a range of triadic coupling strengths $\sigma_\Delta$. 
(c) Zoom-in view around $\lambda_{\max}=0$ showing which twisted states are stable ($\lambda_{\max}<0$). More twisted states become stable as $\sigma_\Delta$ is increased. 
(d) Fraction of twisted states that are stable as a function of $\sigma_\Delta$, for $n=83$ and $n=830$.
}
\label{fig:linear_stability}
\end{figure*}

Here, we focus on an analytically tractable case of \cref{eq:kuramoto} equipped with a simple ring structure: 
\begin{equation}
\begin{split}
    \dot{\theta}_i = & \frac{\sigma}{2r} \sum_{j=i-r}^{i+r}\sin(\theta_j-\theta_i) \\
    & + \frac{\sigma_\Delta}{2r(2r-1)} \sum_{j=i-r}^{i+r} \sum_{k=i-r}^{i+r} \sin(\theta_j+\theta_k-2\theta_i), 
\end{split}
\label{eq:kuramoto_ring}
\end{equation}
where $r$ is the coupling range.
For the triadic coupling, we require $i\neq j\neq k$, so that each triangle involves three distinct nodes.
Notice that we have set $\omega=0$ by going into a rotating frame. 
Also note that we can always set $\sigma=1$ by rescaling time, which we adopt in simulations throughout the paper. 
For demonstration purposes, all numerical results below are presented for \cref{eq:kuramoto_ring} with $n=83$ and $r=2$, unless otherwise stated. 
The key findings remain qualitatively unchanged for other choices of the parameters.
{\revise The choice of $n=83$ simply follows the convention from earlier papers \cite{delabays2017size,zhang2021basins} and $r=2$ is the smallest coupling range that allows nontrivial simplicial complexes (a popular class of hypergraphs that are heavily studied in the literature), which we investigate later in the paper.}

Due to the rotational symmetry, twisted states are {\rev equilibria of} \cref{eq:kuramoto_ring}.
We first show that triadic interactions can stabilize twisted states far beyond what is possible with pairwise coupling.
To analyze the linear stability of any fixed-point solution $\bm{\theta}^*$ of \cref{eq:kuramoto} {\rev (which includes \cref{eq:kuramoto_ring} as a special case)}, we can write down the Jacobian $\bm{J}(\bm{\theta}^*) = \bm{J}^{(1)}(\bm{\theta}^*) + \bm{J}^{(2)}(\bm{\theta}^*)$, where
\begin{equation}
\begin{split}
    J_{ij}^{(1)}\left(\boldsymbol{\theta}^*\right) = & \frac{\sigma}{k_i^{(1)}} A_{i j} \cos \left(\theta_j^*-\theta_i^*\right),\\
    J_{ij}^{(2)}\left(\boldsymbol{\theta}^*\right) = & \frac{\sigma_\Delta}{k_i^{(2)}} \sum_{k=1}^n B_{ijk} \cos \left(\theta_j^*+\theta_k^*-2\theta_i^*\right)
\end{split}
\label{eq:jac}
\end{equation}
for $i\neq j$ and {\rev $J_{ii}^{(\ell)} = -\sum_{\substack{j=1 \\ j\neq i}}^n J_{ij}^{(\ell)}$}.
For the rotationally symmetric topology considered in \cref{eq:kuramoto_ring}, the Jacobian is a symmetric circulant matrix of the form
{\rev
\begin{equation}
\bm{J}=\left[\begin{array}{ccccccc}
J_0 & J_1 & J_2 & \cdots & J_3 & J_2 & J_1 \\
J_1 & J_0 & J_1 & \cdots & J_4 & J_3 & J_2 \\
\vdots & \vdots & \vdots & \ddots & \vdots & \vdots & \vdots \\
J_2 & J_3 & J_4 & \cdots & J_1 & J_0 & J_1 \\
J_1 & J_2 & J_3 & \cdots & J_2 & J_1 & J_0
\end{array}\right].
\end{equation}
}
We know that the normalized eigenvectors of a circulant matrix are the Fourier modes and the eigenvalues of $\bm{J}$ are given by
\begin{equation}
    \lambda_p(\bm{J})=\sum_{s=0}^{n-1} J_s e^{2 \pi \mathrm{i} p s / n}, \quad 0 \leq p \leq n-1,
\label{eq:eigs}
\end{equation}
where $J_s=J_{n-s}$ for $\floor{n/2}<s<n$ \cite{davis1979circulant}.
This implies that $\lambda_p$ are all real. 
Moreover, $\lambda_0$ is always equal to $0$, and it represents the mode for which all oscillators are perturbed by the same amount.
For any $q$-twisted state, $\bm{\theta}^{(q)}$, $J_s$ is given by
\begin{equation*}
\begin{split}
    J_s = & \, \frac{\sigma}{2r} \cos(\frac{2\pi q}{n}s) + \frac{\sigma_\Delta}{r(2r-1)} \sum_{k=-r}^{r} \cos(\frac{2\pi q}{n}(s+k)) \\
    & - \frac{\sigma_\Delta}{r(2r-1)} \sum_{j=1}^{2} \cos(\frac{2\pi q}{n}js) \text{ for } 0< s \leq r, \\
    J_s = & \, 0 \text{ for } s>r , \\
    J_0 = & -2\sum_{s=1}^r J_s.
\end{split}
\label{eq:Js}
\end{equation*}
For example, for $r=2$,
\begin{equation*}
\begin{split}
    J_1 = & \frac{\sigma}{4} \cos(\frac{2\pi q}{n}) + \frac{\sigma_\Delta}{6} \left[ 1 + \cos(\frac{2\pi q}{n}) + \cos(\frac{6\pi q}{n}) \right],\\
    J_2 = & \frac{\sigma}{4} \cos(\frac{4\pi q}{n}) + \frac{\sigma_\Delta}{6} \left[ 1 + \cos(\frac{2\pi q}{n}) + \cos(\frac{6\pi q}{n}) \right].
\end{split}
\label{eq:r=2}
\end{equation*}
Plugging the formula for $J_s$ into \cref{eq:eigs}, we can analytically obtain the spectrum of the Jacobian for any $q$-twisted state and any coupling range $r$.

In \cref{fig:linear_stability}(b)-(c), we show the linear stability of twisted states, measured by {\revise the largest Lyapunov exponent transverse to the synchronization manifold,} $\lambda_{\max}=\max\{\lambda_1,\lambda_2,\dots,\lambda_{n-1}\}$, as a function of the winding number $q$.
{\revise First, note that we show $n$ values of $q$ because, by definition, there are only $n$ distinct twisted states: $q \to q + n $ simply adds $2\pi$ to all phases, leaving them unchanged. 
We consider $- \floor{\frac{n}{2}} \leq q \leq \floor{\frac{n}{2}}$~\footnote{Because twisted states with winding numbers $-q$ and $n-q$ states are the same, one could equivalently consider the interval $ 0 \leq q \leq n$.}.
Second, note that the plot is symmetric with respect to $q=0$. 
This follows from the fact that twisted states with winding numbers $q$ and $-q$ are the same up to the reversal symmetry $\bm{\theta} \rightarrow -\bm{\theta}$.
Because of this symmetry, below we only need to describe states with positive $q$.}
At $\sigma_\Delta=0$, the stability curve has three peaks, and $\lambda_{\max}$ becomes positive for $q>11$.
Thus, pairwise coupling in \cref{eq:kuramoto_ring} cannot support stable twisted states with more than $11$ twists (for $n=83$ and $r=2$).
For systems with strong triadic couplings, two of the peaks are flattened.
As a result, a lot more twisted states become stable.
For example, all twisted states up to $33$ twists are stable for $\sigma_\Delta=4$.
Interestingly, for intermediate $\sigma_\Delta$, because $\lambda_{\max}$ is a nonmonotonic function of $q$, winding numbers on disjoint intervals can become stabilized.
For instance, at $\sigma_\Delta=1$, twisted states are stable for $q\leq 14$, become unstable for $14<q<23$, then become stable again for $23\leq q \leq 30$ (see \cref{fig:linear_stability}(c) for details).
{\rev Another interesting phenomenon to note is that the gradient of $\lambda_{\max}$ becomes much steeper upon crossing the $x$-axis from below.
This is caused by the switch of the dominant eigenvector, which we further elobrate in \cref{fig:lambda2_crossing}.}

\Cref{fig:linear_stability}(d) shows the fraction of stable twisted states $P_\text{stable}$ as a function of $\sigma_\Delta$, which further emphasizes the dramatic number of twisted states stabilized by triadic interactions.
We see that $P_\text{stable}$ is monotonically increasing with $\sigma_\Delta$, and 
that one can easily go from less than $30\%$ of stable twisted states to over $80\%$ stable by adding triadic interactions.
We also show the same curve for a larger system with $n=830$, which is basically a smoother version of $n=83$ (since there are a lot more twisted states for $n=830$).
Our results echo the recent findings in Ref.~\cite{bick2023phase}, which showed that higher-order interactions can stabilize twisted states in graphons.

\subsection{Basin stability analysis}

\begin{figure}[b]
\centering
\includegraphics[width=1\columnwidth]{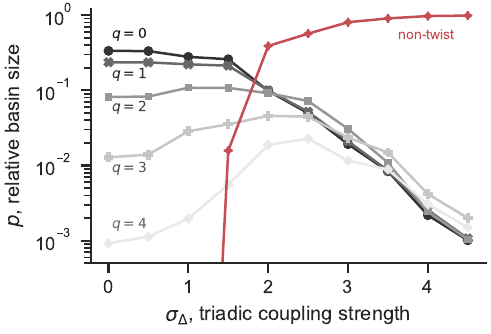}
\caption{
\textbf{Higher-order interactions decrease the basin stability of twisted states.} 
We show $p$, the relative basin size, as a function of the triadic coupling strength $\sigma_\Delta$.
We estimated $p$ by simulations starting from $10^5$ random initial conditions.
Each line represents a $q$-twisted state, except the red line, which represents attractors that are not twisted states. The relative basin size of non-twisted states quickly approaches $1$ as $\sigma_\Delta$ is increased.
We only show $q\geq 0$ due to the symmetry between $q$ and $-q$.
}
\label{fig:basin_stability}
\end{figure}

Next, we switch to the basin stability perspective.
\Cref{fig:basin_stability} shows how the relative basin size $p$ of the twisted states changes with the triadic coupling strength $\sigma_\Delta$. 
{\rev We compute the basin size relative to the full state space (which is compact) by simulating the dynamics starting from $10^5$ random initial conditions and counting the proportion of those that converge to a given state.}
For $\sigma_\Delta=0$, twisted states (including full synchrony), are the only stable states and thus take up the entire state space \cite{diaz2024exploring} {\revise (see Supplementary Materials Section S1 for details)}. 
States with fewer twists (smaller $q$) have a larger basin size. 
In particular, full synchrony attracts the most initial conditions.
Now, for small $\sigma_\Delta$, triadic interactions are affecting twisted states unequally: the basins for small $q$ shrink whereas those for large $q$ expand.  
As $\sigma_\Delta$ is further increased, the basins for the non-twisted states appear and quickly become dominant, whereas the basins of the twisted states all shrink and become comparable in size. 
Note that among the twisted states, full synchrony does not have the largest basin anymore---in fact, the twisted state with the largest basin has more twists as $\sigma_\Delta$ is increased.

\begin{figure*}[htb]
\centering
\includegraphics[width=1\linewidth]{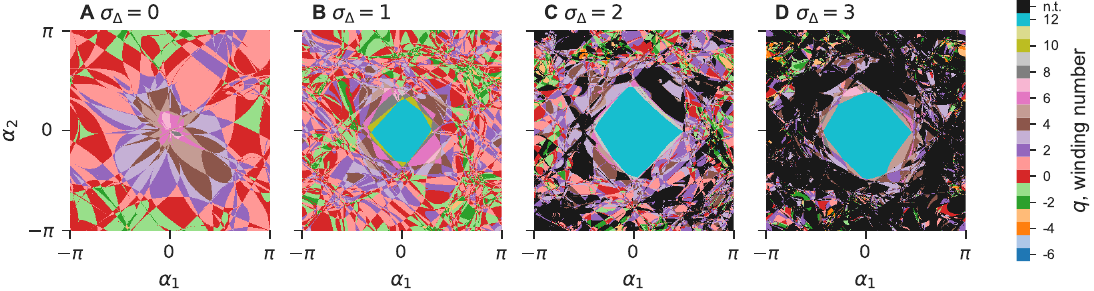}
\caption{
\textbf{Higher-order interactions stabilize twisted states but shrink their basins.} 
Two-dimensional slices of the state space (centered around the twisted state with $q=12$) showing how basins change as $\sigma_\Delta$ is increased. 
The basins of the twisted states are colored according to their winding number $q$, and the basins of all other states are colored black. 
(a)~For $\sigma_\Delta=0$, the twisted state with $q=12$ is unstable, and all points converge to a twisted state with a lower winding number.
(b)~For $\sigma_\Delta=1$, all attractors are still twisted states.
Moreover, $q=12$ becomes stable, which creates the cyan basin at the center.
(c, d)~For stronger triadic interactions ($\sigma_\Delta=2$ and $3)$, the $q=12$ basin expands, but non-twisted states also start to appear and quickly become dominant.
Although the basin for $q=12$ looks significant on the 2D slices, due to the high-dimensional nature of the state space, it would be almost impossible to reach from random initial conditions.
}
\label{fig:basin_slice}
\end{figure*}

\Cref{fig:basin_slice} further illustrates the opposite effects of higher-order interactions on linear stability and basin stability by visualizing the morphology of basins as $\sigma_\Delta$ is increased.
Specifically, we examine a random 2D slice of the state space, spanned by $\bm{\theta}_0 + \alpha_1\bm{P}_1 + \alpha_2\bm{P}_2$, $\alpha_i \in (-\pi,\pi]$.
Here, $\bm{P}_1$ and $\bm{P}_2$ are $n$-dimensional binary orientation vectors in which $\floor*{n/2}$ randomly selected components are $1$ and the rest of the components are $0$.
We set the origin to be the twisted state with $q=12$, $\bm{\theta}_0 = \bm{\theta}^{(12)}$, which we know from \cref{fig:linear_stability} is unstable when $\sigma_\Delta=0$.
Thus, we can only see basins for $q$ between $-3$ and $7$ in \cref{fig:basin_slice}(a).
Adding triadic interactions stabilizes $q=12$, so we see its basin emerge in \cref{fig:basin_slice}(b) when $\sigma_\Delta$ is set to~$1$.
For larger $\sigma_\Delta$ shown in panels (c) and (d), the reduction in basin stability for twisted states becomes apparent.
Despite their substantially improved linear stability, the basins for twisted states (as a group) rapidly shrink as more and more points get absorbed into the basin for non-twisted states (colored black), making twisted states hard to reach from random initial conditions.

\begin{figure}[b]
\centering
\includegraphics[width=0.9\columnwidth]{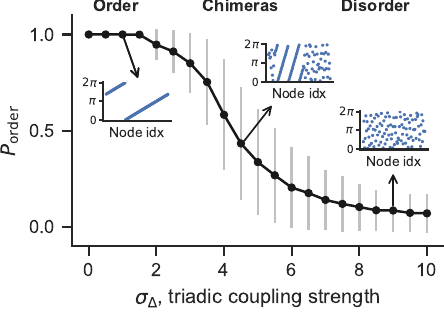}
\caption{
\textbf{Chimeras bridge the transition from order to disorder as triadic interactions become stronger.} 
As we increase $\sigma_\Delta$, there is a smooth transition from order (twisted states) to chimeras, and then to disorder.
We monitor the transition by computing the portion of ordered oscillators $P_\text{order}$ in the final state, averaged over $1000$ simulations from random initial conditions for each $\sigma_\Delta$.
The error bars represent standard deviations and the insets show typical attractors reached from random initial conditions.
}
\label{fig:chimeras}
\end{figure}

The natural next question is: What are those non-twisted states created by higher-order interactions?
\Cref{fig:chimeras} shows that they consist of chimera states with increasingly large disordered domains as triadic couplings become stronger.
Here, $P_\text{order}$ measures the portion of oscillators that are ordered.
Twisted states correspond to $P_\text{order}=1$ and disordered states correspond to $P_\text{order}\approx0$, whereas chimera states have $P_\text{order}$ between $0$ and $1$.
For final states reached from random initial conditions, the average $P_\text{order}$ decreases gradually from $1$ to $0$ as $\sigma_\Delta$ is increased.

We classify oscillators as (dis)ordered by calculating the local order parameters 
\begin{equation}
    O_j = \frac{1}{2r+1}\sum_{k=j-r}^{j+r} e^{\mathrm{i}\theta_k},
\end{equation}
for $j = 1, \dots, n$.
We classify oscillator $j$ as disordered if $|O_j|<0.85$.
The value of 0.85 is large enough to ensure $P_\text{order} \simeq 0$ for large $\sigma_\Delta$, which matches visual inspection of the states.
{\revise Although this analysis is phenomenological in nature,} the twisted states would have $|O_j|\approx 1$ if the winding numbers and coupling ranges are small, which are satisfied for the systems in \cref{fig:chimeras}.

\begin{figure*}[ht]
\centering
\includegraphics[width=0.95\linewidth]{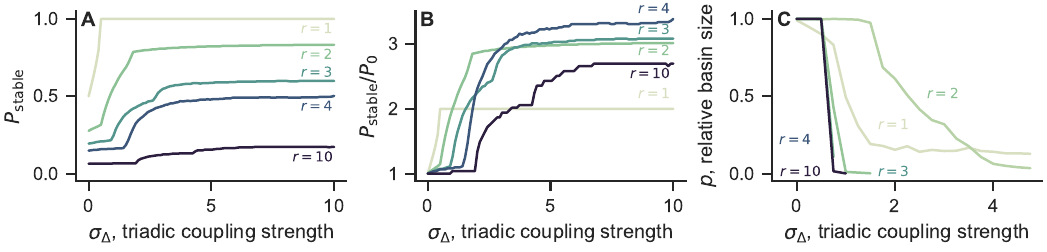}
\caption{
\textbf{Higher-order interactions increase linear stability while decreasing basin stability for hypergraph rings with different coupling ranges.} 
(a) Fraction of twisted states that are stable as a function of $\sigma_\Delta$, calculated with $n=830$.
(b) Same data as in (a), but shown as the ratio between the number of stable twisted states at nonzero $\sigma_\Delta$ and $\sigma_\Delta=0$.
(c)~Relative basin size of all twisted states combined as a function of $\sigma_\Delta$. For any given $r$, twisted states become more difficult to find as the triadic couplings become stronger. The basin sizes are estimated by simulating $n=83$ oscillators from $10^3$ random initial conditions.
}
\label{fig:HG_r}
\end{figure*}

The insets show typical attractors for different values of $\sigma_\Delta$.
Despite the fact that more and more twisted states become linearly stable for larger $\sigma_\Delta$, they are increasingly unlikely to be observed from random initial conditions.
Instead, the state space is dominated by the basins for chimera states (intermediate $\sigma_\Delta$) or disordered states (large $\sigma_\Delta$).
This is consistent with recent results showing that higher-order interactions promote chimera states in simplicial complexes \cite{kundu2022higher}.
{\revise Similar states have also been observed recently in a continuous-space system~\cite{smirnov2024nonuniformly}.}
We note that the exact appearance of chimeras or disordered states can vary for different coupling ranges or coupling structures.
However, the order-chimera-disorder transition described here is a robust phenomenon.

{\rev We also note that here the disorder is only in space, not in time---the patterns either remain frozen over time (fixed points) or rotate uniformly with a constant speed (periodic orbits). 
In particular, all oscillators are phase-locked.
This is in contrast to traditional chimeras, for which the disordered oscillators are not frequency synchronized and their relative phases change over time.
In \cref{fig:chimera_feq}, we show the statistics of the {\revise effective} frequency for different initial conditions and under different coupling strengths $\sigma_\Delta$.
For small $\sigma_\Delta$, the {\revise effective} frequency is always $0$, but nonzero frequency can emerge for $\sigma_\Delta \ge 2$.
This is consistent with recent results showing that twisted states can undergo Hopf bifurcations as $\sigma_\Delta$ is increased \cite{bick2023hopf}, {\revise which also consider the effect of higher-order interactions on Kuramoto oscillators but differ from our systems in crucial details}.
}

\subsection{Other coupling structures}


The above results for $r=2$ remain qualitatively unchanged for other coupling ranges $r$.
\Cref{fig:HG_r} shows the opposite behavior of linear and basin stability for a wide range of $r$.
For $r=1$, $50\%$ of the twisted states are stable under purely pairwise coupling.
As $\sigma_\Delta$ is increased, all twisted states quickly become stabilized.
As the coupling structure becomes more nonlocal (larger $r$), fewer twisted states are stable at $\sigma_\Delta=0$.
By introducing triadic couplings, one can always have at least twice as many stable twisted states. For $r \ge 3$, we also observed the appearance of 2-cluster states, that is, with oscillators split (potentially unequally) into two $\pi$-separated clusters (described, e.g., in \cite{golomb1992clustering, skardal2019abrupt,gong2019lowdimensional}), as shown in \cref{fig:SM_coupling range,fig:SM2_coupling range}.


Aside from the generalization of Kuramoto models introduced in \cref{eq:kuramoto_ring}, there are several other natural ways to introduce triadic interactions. 
For example, we can turn the (pairwise) ring network into a simplicial (flag) complex by filling all pairwise triangles.
This implies that $B_{ijk}=1$ if and only if $i\neq j\neq k$, and all three pairs are within distance $r$: $|(i-j) \mod n|\leq r$, $|(i-k) \mod n|\leq r$, and $|(j-k) \mod n|\leq r$.
In other words, $B_{ijk}=A_{ij}A_{ik}A_{jk}$.
In comparison, the topologies considered in \cref{eq:kuramoto_ring} do not require $|(j-k) \mod n|\leq r$ for $B_{ijk}=1$.
This can be expressed equivalently as $B_{ijk}=A_{ij}A_{ik}$.
As shown in \cref{fig:SC}, the main results presented above for ring hypergraphs remain valid for simplicial complexes.


\begin{figure}[b]
\centering
\includegraphics[width=0.9\columnwidth]{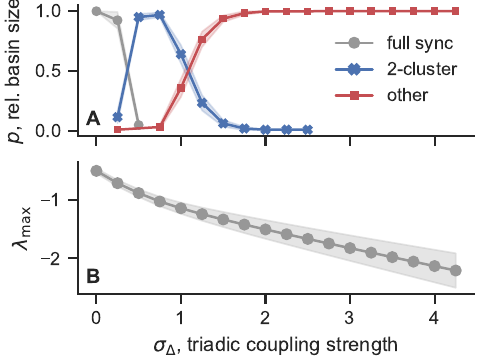}
\caption{
\textbf{Synchronization becomes linearly more stable but harder to reach in random hypergraphs}. We show (a) the relative basin size of full synchrony (0-twisted), 2-cluster, and other (non-twisted) states, and (b) the maximum transverse Lyapunov exponent for full synchrony. Results show the average over 20 random hypergraph realizations {\rev with $n=83$ nodes} (with 100 initial conditions each). The shaded areas indicate one standard deviation.
}
\label{fig:randomHG}
\end{figure}

Finally, we demonstrate that a similar phenomenon persists in more irregular structures using random hypergraphs. 
{\rev A random hypergraph is a generalization of Erdős–Rényi random graphs: We add a hyperedge between any two nodes with probablity $p_1$ and between any three nodes with probability $p_2$. We set $p_d = 20 / n^d$, for $d=1, 2$.}
For random hypergraphs, the only twisted state that can be an attractor is full synchrony ($q=0$). 
Similar to what we found above for more regular structures, in random hypergraphs triadic interactions make full synchrony linearly more stable \cite{lucas2020multiorder,zhang2023higher} but its basin of attraction shrinks dramatically in favor of 2-cluster states---for $\sigma_{\Delta}$ up to around 1.5---and then more disordered states for stronger triadic coupling strengths (\cref{fig:randomHG}). 


\section{Discussion}

In this paper, we showed that higher-order interactions can make basins deeper but smaller---attractors become linearly more stable but at the same time are harder to find due to their basins shrinking dramatically.
We demonstrated this phenomenon for Kuramoto dynamics with 
a wide range of coupling structures (ring hypergraphs with different coupling ranges, ring simplicial complexes, and random hypergraphs).
We were able to characterize the linear stability of all twisted states analytically under these 
coupling structures.
For basin stability, our systematic numerical simulations revealed interesting global features of the dynamics as $\sigma_\Delta$ is increased.
In particular, the basins of twisted states become deeper but smaller due to the proliferation of new attracting states. 
We further characterized these new states introduced by higher-order interactions that compete with twisted states, which manifest as 2-cluster, chimera, or disordered states depending on the ratio between $\sigma$ and $\sigma_\Delta$. 

Deeper but smaller basins induced by higher-order interactions can confer functional advantages to some biological systems.
For instance, for the brain to function optimally \cite{khona2022attractor}, the attractors should have high linear stability so the brain can quickly return to the current state when subject to small perturbations or noise. 
At the same time, we also want the brain to be nimble and able to transition among different states efficiently (e.g., during computation and information processing), which can be achieved by having small basins.

Why do the basins of twisted states shrink as higher-order couplings are introduced?
First, we note that, unlike their pairwise counterparts, Kuramoto systems with nonpairwise interactions 
are generally not gradient systems \cite{bick2023hopf}.
This provides the freedom for \cref{eq:kuramoto,eq:kuramoto_ring} to undergo Hopf bifurcations as $\sigma_\Delta$ is increased.
Böttche et al.~\cite{bottcher2023local} showed recently that anti-correlation between linear stability and basin stability often emerges for dynamical systems that undergo consecutive Hopf bifurcations, offering a potential mechanism for higher-order interactions to shrink basins.
More generally, {\new in a compact phase space,} as more attractors are created, the average basin size would decrease.
In our case, the new states that emerge are more disordered than twisted states and they hold enormous entropic advantages (there are many more possible disordered configurations than ordered ones).
Even 2-cluster states, which appear ordered on the surface, have many more configurations than twisted states---the oscillators can be divided between the two clusters in $2^n$ different ways \cite{skardal2019abrupt}. 

Does extensive multistability emerge naturally from generic higher-order interactions regardless of details about the dynamics and couplings?
Such phenomena have been observed under many different settings \cite{tanaka2011multistable,skardal2019abrupt,kundu2022higher,skardal2023multistability}.
For \cref{eq:kuramoto} with all-to-all coupling, it was shown previously that the coupling function $\sin(\theta_j+\theta_k-2\theta_i)$ introduces a higher-order harmonic and nonlinear dependence on the order parameter in the mean-field description, which create additional nonlinearity and extensive multistability in the self-consistent equations for the order parameters \cite{xu2020bifurcation,xu2021spectrum}. 
The same is true for a different coupling function, $\sin(2\theta_j-\theta_k-\theta_i)$:
Using the Ott-Antonsen ansatz \cite{ott2008low}, it was found that higher-order interactions give rise to added nonlinearity in the reduced equations that describe the macroscopic system dynamics \cite{skardal2020higher}. 
{\new Finally, when deviating from all-to-all coupling (such as the local couplings considered here)}, we expect the nontrivial coupling structure could introduce additional nonlinearity into the macroscopic equations, further increasing multistability.

{\new It is possible that deeper but smaller basins may not be the exclusive results of higher-order interactions. Can other forms of nonlinear coupling functions create similar effects? If so, what are the properties required of those nonlinearities?}
{\revise For example, another way to add nonlinearity to the Kuramoto model is through higher-harmonic coupling functions such as $\sin(2\theta_j - 2\theta_i)$ \cite{suda2015persistent,gong2019lowdimensional}.
In our preliminary tests, however, we found that they do not stabilize more twisted states.}
{\new The systematic exploration of the link between nonlinearity and stability (both local and global) is an important and open question, which we leave for future works.}

In conclusion, the prevalence of anti-correlation between linear stability and basin stability warrants a more nuanced and comprehensive approach when considering collective dynamics on hypergraphs and simplicial complexes. 
Understanding the global organization of attractors and saddles in the presence of nonpairwise couplings is crucial to the prediction and control of complex systems such as ecological communities and neuronal populations.
We hope this work will stimulate future endeavors to understand the effects of higher-order interactions from both local and global perspectives.





\bibliography{bibli}



\vspace{10mm}

\noindent \textbf{Acknowledgements:} 
%
We thank Melvyn Tyloo, Arkady Pikovsky, and Michael Rosenblum for insightful discussions.\\
\noindent \textbf{Funding:} YZ acknowledges support from the Omidyar Fellowship.\\
\noindent \textbf{Author Contributions:} YZ and ML designed the research. YZ, PSS, and ML performed analysis and simulations. All authors discussed the results. YZ and ML wrote the paper with feedback from PSS, FB, and GP.\\
\noindent \textbf{Data and materials availability:} 
Code for reproducing our results is available online from the repository \url{https://github.com/maximelucas/basins_and_triangles} and the Zenodo reference \url{https://zenodo.org/doi/10.5281/zenodo.11105328}. It utilizes the XGI package \cite{Landry_XGI_2023}.

\clearpage

\newcommand\SupplementaryMaterials{%
  \xdef\presupfigures{\arabic{figure}}
  \xdef\presupsections{\arabic{section}}
  \renewcommand{\figurename}{Supplementary Figure}
  \renewcommand\thefigure{S\fpeval{\arabic{figure}-\presupfigures}}
  \renewcommand\thesection{S\fpeval{\arabic{section}-\presupsections}}
}

\SupplementaryMaterials

\clearpage
\onecolumngrid
\setcounter{page}{1}
\renewcommand{\thepage}{S\arabic{page}}
\setcounter{equation}{0}
\renewcommand{\theequation}{S\arabic{equation}}

\begin{center}
{\Large\bf Supplementary Materials}\\[5mm]
{\large{Deeper but smaller: Higher-order interactions increase linear stability but shrink basins}}\\[5pt]
Yuanzhao Zhang, Per Sebastian Skardal, Federico Battiston, Giovanni Petri, and Maxime Lucas
\end{center}


{\revise
\section{Global stability of twisted states on ring networks}

In pairwise ring networks of identical Kuramoto oscillators, all attractors are twisted states.
The basic arguments are as follows: Identical Kuramoto oscillators on symmetric networks are described by the gradient of a smooth potential on the $n$-torus. 
Because gradient dynamics on a compact manifold can only have fixed points as attractors, we can rule out periodic or chaotic solutions.
For fixed points, because of the ring structure, the steady-state phase differences between neighboring oscillators have to be either $\delta$ or $\pi-\delta$, with a $\delta = 2\pi q/n$ that is independent of the node indices. 
Through linear stability analysis (or simply by using the Gershgorin circle theorem), one can show that the only fixed points that are stable are those with all phase differences $\theta_{i+1}-\theta_i$ equal to $\delta$.
This then establishes that twisted states are the only attractors.
}

{\rev
\section{Crossing of eigenmodes}

In \cref{fig:linear_stability}, the steepness of the $\lambda_{\max}$ curves changed dramatically at the instability transition because the dominant eigenvector of the Jacobian changed. The basic intuition is that there are eigenmodes that oscillate much more ``wildly'' than others. In \cref{fig:lambda2_crossing} below, each curve represents an eigenvalue. The orange curve is the most ``tame'' curve and usually gives the largest transverse Lyapunov exponent $\lambda_{\max}$ when all the eigenvalues are negative. However, around $\lambda=0$, it gets overtaken by other eigenmodes with much steeper gradients. This explains why $\lambda_{\max}$ suddenly shoots up rapidly after crossing the $x$-axis from below.

\begin{figure}[h]
\centering
\includegraphics[width=.45\columnwidth]{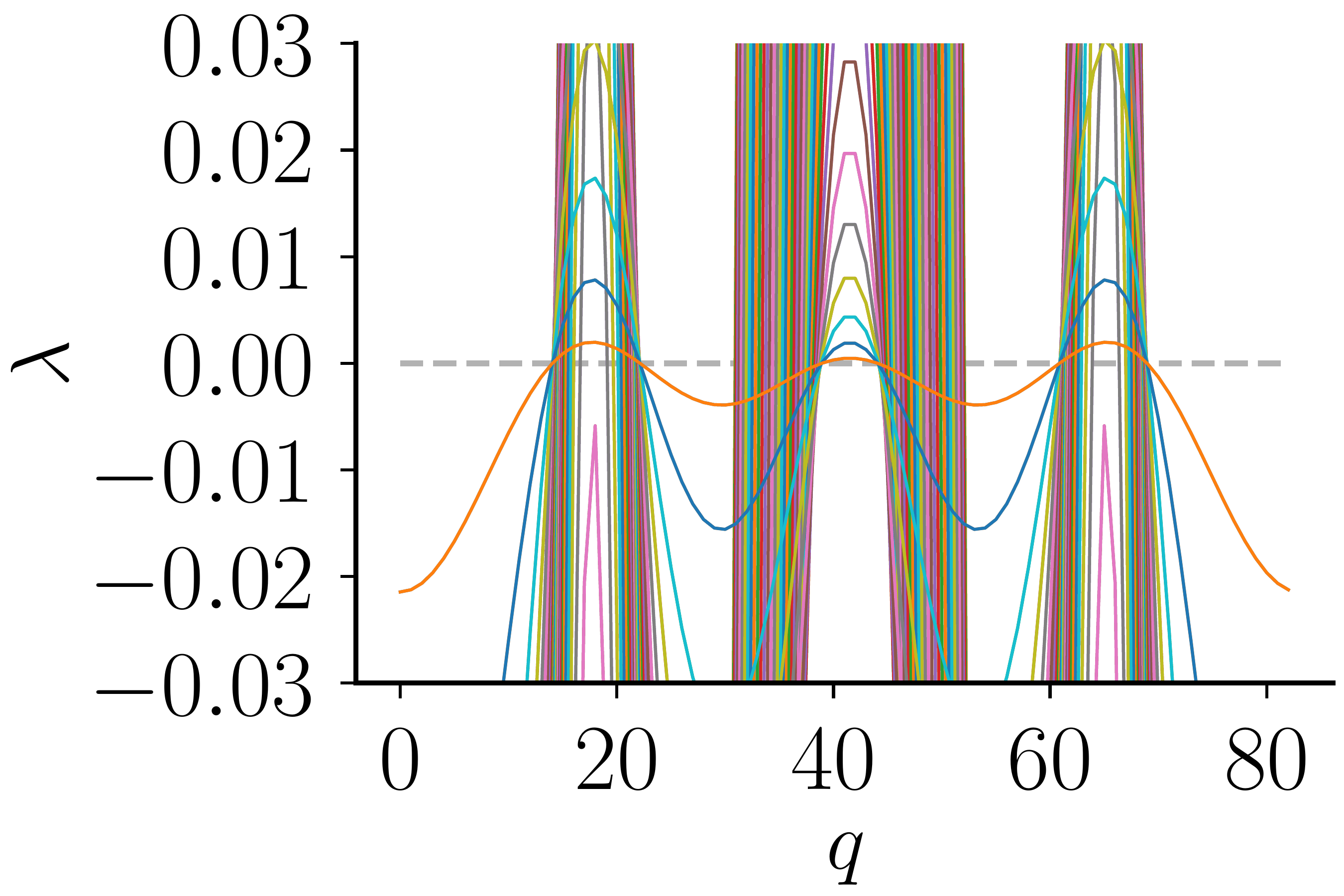}
\includegraphics[width=.45\columnwidth]{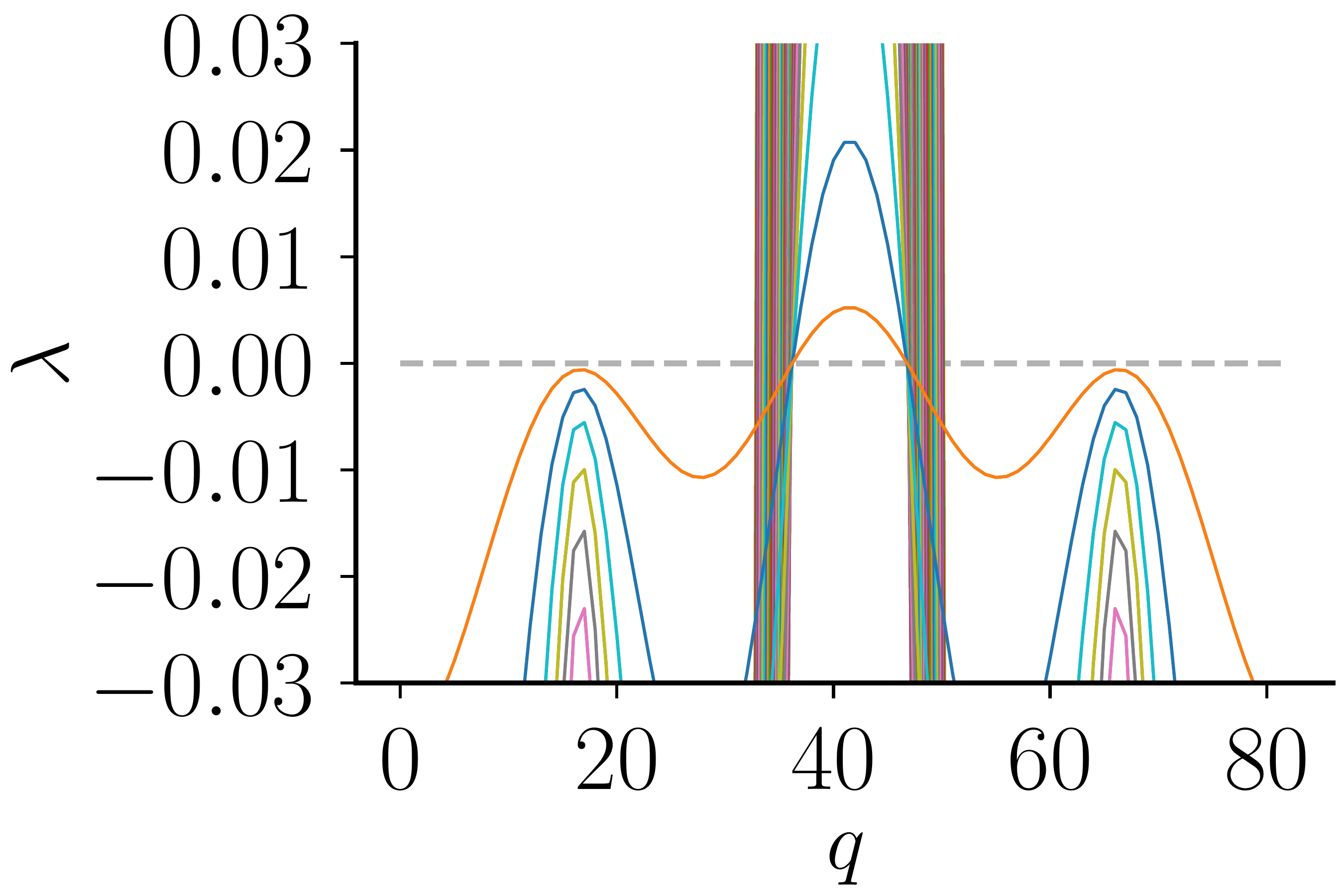}
\vspace{-4mm}
\caption{\rev Full spectrum of the Jacobian for the system in Fig.~1 with $\sigma_\Delta = 1$ (left) and $\sigma_\Delta = 2$ (right).}
\label{fig:lambda2_crossing}
\end{figure}
}

\newpage

{\rev
\section{Frequency analysis of chimera states}

For the states shown in \cref{fig:chimeras}, the rotation speed is always constant and identical across all oscillators, regardless of whether they are spatially ordered, disordered, or a mixture of both.
\Cref{fig:chimera_feq} shows the {\revise effective} frequency $\Omega$ for different initial conditions and under different coupling strengths $\sigma_\Delta$.
Here, $\Omega$ is computed as an average over all oscillators and over time.
However, this is not crucial as all oscillators have the same frequency and are moving with constant speed.

\begin{figure}[h]
\centering
\includegraphics[width=.43\columnwidth]{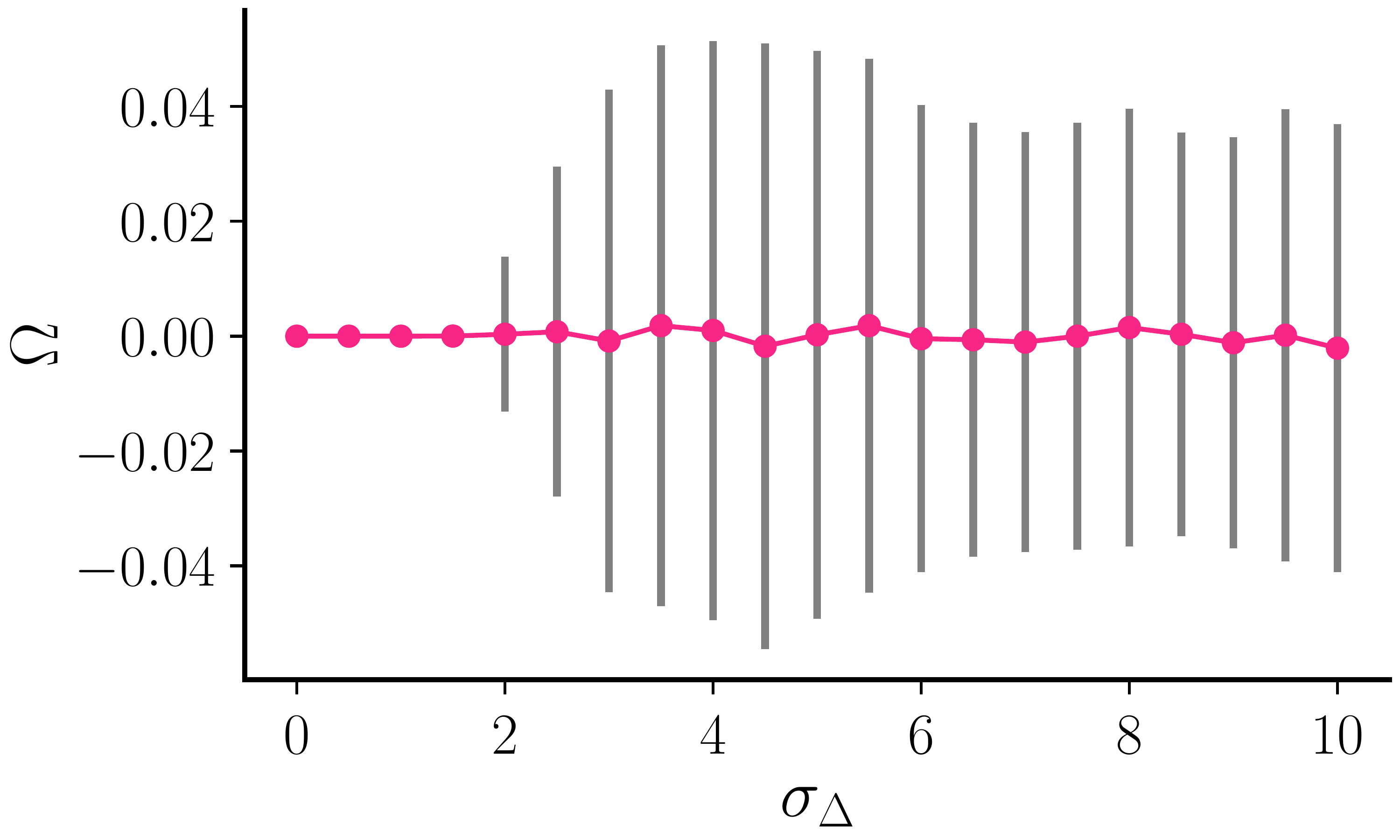}
\includegraphics[width=.43\columnwidth]{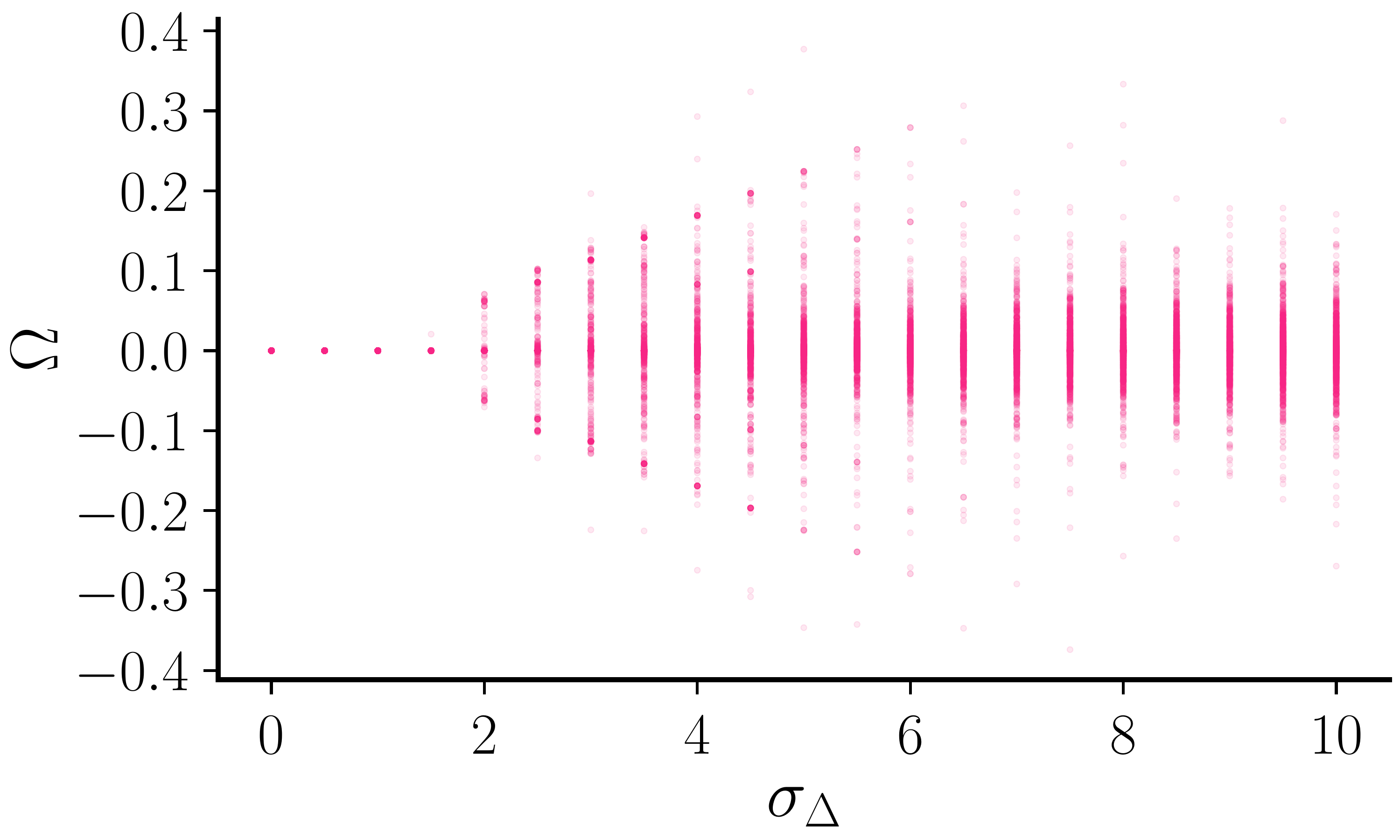}
\vspace{-4mm}
\caption{\rev {\revise Effective} frequency $\Omega$ of the oscillators for \cref{eq:kuramoto_ring} with $n=83$ and $r=2$. At each value of $\sigma_\Delta$, $1000$ initial conditions were simulated. The left panel shows the average over initial conditions and the corresponding standard deviations. The right panel shows a scatter plot of the {\revise effective} frequency for each initial condition.}
\label{fig:chimera_feq}
\end{figure}
}

\section{Ring hypergraphs with different coupling ranges}

\begin{figure}[h!]
\centering
\includegraphics[width=0.29\linewidth]{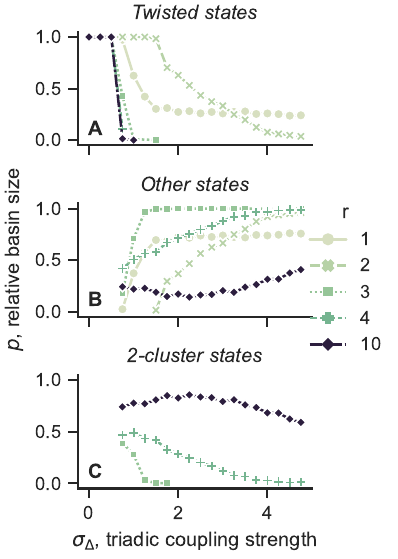}
\caption{
\textbf{Extension of \cref{fig:HG_r}c for ring hypergraphs with different coupling ranges $r$.} 
Here we also show relative basin sizes of {\rev (A) twisted states, (B) other states, and (C) 2-cluster states}. 
Two-cluster states appear for $r \ge 3$.
}
\label{fig:SM_coupling range}
\end{figure}

\begin{figure}[h!]
\centering
\includegraphics[width=0.23\linewidth]{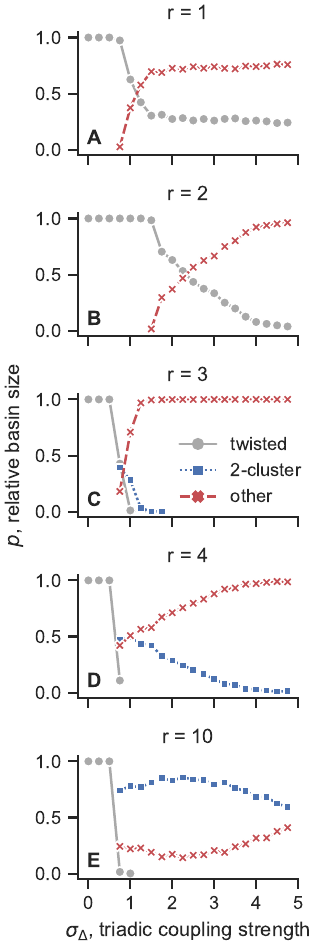}
\caption{
\textbf{Same data as in \cref{fig:SM_coupling range} for ring hypergraphs with different couping ranges.} 
Here we show the basin sizes per value of $r$.
}
\label{fig:SM2_coupling range}
\end{figure}

{\rev Here, we consider the dynamics on ring hypergraphs from \cref{eq:kuramoto_ring} and vary the coupling range $r$. In \cref{fig:SM_coupling range}, we show the relative basin sizes of the twisted states, the 2-cluster states, and the other states remaining. Here, we are mainly interested in how much space each category of states takes. Therefore, the basin size for the twisted states is aggregated over all twisted states, 
and that of the 2-cluster states is aggregated over all possible configurations of the two clusters. 
Consistent with our results in the main text, the basin size associated with twisted states drops for larger triadic coupling strengths while the basin associated with other states takes most of the space. An additional type of state also appears---as described in the main text, in these 2-cluster states, oscillators are split (usually unequally) into two $\pi$-separated clusters. 
\Cref{fig:SM2_coupling range} shows the same data differently: Each panel corresponds to a single value of $r$ and shows one curve per category of states, more similarly to \cref{fig:basin_stability}.}

\newpage

\section{Ring simplicial complexes}

For simplicial complexes, we consider systems described by the following equations:
\begin{equation}
    \dot{\theta}_i = \frac{\sigma}{2r} \sum_{j=i-r}^{i+r}\sin(\theta_j-\theta_i) + \frac{\sigma_\Delta}{3r(r-1)} \sum_{\substack{0<|k-i|\leq r \\ 0<|j-i|\leq r \\ 0<|j-k|\leq r}} \sin(\theta_j+\theta_k-2\theta_i), \quad i=1,\dots,n.
\end{equation}
It is easy to verify that the coupling structure forms a simplicial complex for any coupling range $r$.

For simplicial complexes with coupling range $r$, we have
\begin{equation}
    J_s = \frac{\sigma}{2r} \cos(\frac{2\pi q}{n}s) + \frac{2\sigma_\Delta}{3r(r-1)} \sum_{k=s-r}^{r} \cos(\frac{2\pi q}{n}(s+k)) - \frac{2\sigma_\Delta}{3r(r-1)} \sum_{j=1}^{2} \cos(\frac{2\pi q}{n}js)
\end{equation}
for $0< s \leq r$.
Specifically, for $r=2$ we have
\begin{equation}
\begin{split}
J_1 = & \frac{\sigma}{4} \cos(\frac{2\pi q}{n}) + \frac{\sigma_\Delta}{3} \left[ 1 + \cos(\frac{6\pi q}{n}) \right],\\
J_2 = & \frac{\sigma}{4} \cos(\frac{4\pi q}{n}) + \frac{\sigma_\Delta}{3} \cos(\frac{6\pi q}{n}).
\end{split}
\end{equation}

\begin{figure*}[htb]
\centering
\includegraphics[width=0.99\columnwidth]{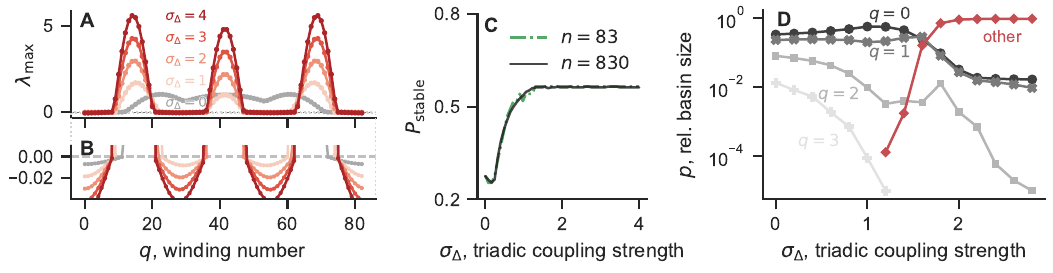}
\caption{
\textbf{Analog of \cref{fig:linear_stability,fig:basin_stability} for ring simplicial complexes.} 
Here, we show detailed results for the coupling range $r=2$, but they remain qualitatively unchanged for larger $r$.
}
\label{fig:SC}
\end{figure*}

We show the results for this structure in \cref{fig:SC}.

\end{document}